\documentclass[prb,preprint,showpacs]{revtex4}
\usepackage{epsfig}
\usepackage{amsmath}
\usepackage{amssymb}
\usepackage{times}
\usepackage{graphicx}
\begin{document}

\title{Quantum hydrodynamic modeling  of edge modes in chiral Berry plasmons}

\author{Ya Zhang$^1$, Feng Zhai$^2$, Bin Guo$^1$$^{*}$, Lin Yi$^{3}$, Wei Jiang$^3$$^{*}$}
\address{$^{1}$ Department of Physics, Wuhan University of Technology, Wuhan 430070, China}
\address{$^{2}$ Department of Physics, Zhejiang Normal University, Zhejiang 321004, China}
\address{$^{3}$ School of Physics, Huazhong University of Science and Technology, Wuhan 430074, China}
\email{binguo@whut.edu.cn}
\email{weijiang@hust.edu.cn}

\begin{abstract}
A  quantum hydrodynamic model is used to study the edge modes of chiral Berry plasmons.
The transcendental equation of the dispersion relation is solved nonlinearly and semi-analytically. We predict a new one-way chiral edge state with the quantum effect compared to that without the quantum effect, at the both side of $q=0$. Indeed, the plasmon frequencies for positive and negative $q$, exhibit different limits for $q\rightarrow 0^{-}$ and $q\rightarrow 0^{+}$.
As a result, the quantum effect enhances the chirality in the vicinity of $q=0$. Both counterpropagating edge modes exhibit greater confinement to the edge with the quantum effect. In addition, new localized edge modes are found with increased Berry flux in both cases, i.e., without and with the quantum effect.
\end{abstract}

\maketitle
The anomalous Hall effect was first discovered by Hall in 1881, and occurs as a result of broken time-reversal symmetry in metallic
ferromagnets with strong spin-orbit coupling\cite{Nagaosa2010}. The time-reversal
symmetry can be broken by the inclusion of an appropriate weak magnetic field\cite{Zhai2005}.
In the early stages of research on this phenomenon, "anomalous behavior" was reported by Karplus and Luttinger\cite{Karplus1954}, which naturally arose in the first microscopic theory of the anomalous Hall effect.
The Berry curvature $\Omega_{n}(\mathbf{k})$  concept was adopted theoretically to relate the anomalous Hall effect to the topological nature of the Hall currents. As a result of the topological Berry phase, the intrinsic anomalous Hall effect can be expressed as an integral over the Fermi surface\cite{Haldane2004}.
Further, in ferromagnets, electrons have an "anomalous velocity" perpendicular to the electric field due to their $\Omega_{n}(\mathbf{k})$. In previous studies, Culcer et al.\cite{Culcer2003} have investigated the anomalous Hall effect in paramagnetic two-dimensional
systems with a external magnetic field, while Wang et al.\cite{Wang2014} have investigated the quantum anomalous
Hall effect in conventional diluted magnetic semiconductors with magnetically doped InAs/GaSb
quantum wells based on the Kane model.
Relevant background information and the basic concept of $\Omega_{n}(\mathbf{k})$ are presented in the review compiled by Xiao et al. \cite{Xiao2010}.

All of the studies mentioned above were performed magnetically. However, the Berry phase can
manifest in various magnetic and nonmagnetic materials. As regards cases without a magnetic field, the Berry phase manifests in gapped Dirac materials, where the time reversal asymmetry is realized via a nonequilibrium
valley polarization\cite{Mak2014,Lensky2015}. Thus, gapped Dirac materials with several valleys, such as
graphene and transition metal dichalcogenide monolayers, have been investigated by Xiao et al.\cite{Xiao2007,Xiao2012}. The authors found a valley dependent Berry phase effect which can
yield a valley contrasting Hall transport by an electric mean\cite{Xiao2007}.
Further, Kumar et al.\cite{Kumar2016} have observed chirality in bulk and edge plasmons
without an external magnetic field, where the valley imbalance induces a net $\Omega_{n}(\mathbf{k})$ (Berry flux $F$).
Note that $F$ can cause chiral plasmons
in the absence of a magnetic field\cite{Kumar2016,Song2016}. Finally, Song et al. have confirmed the existence of counterpropagating charge density waves in chiral plasmons, where splitting dispersion occurs
for transverse edge modes in opposite directions. These modes are regarded as chiral Berry plasmons (CBPs)\cite{Song2016}.

However, quantum effects may yield other interesting phenomena in these CBPs.
In this Letter, we employ a quantum hydrodynamic (QHD) method coupled with the anomalous velocity to examine the quantum effect on the splitting dispersive relation and the transverse mode confinement of CBPs propagating along the boundary of a two-dimensional electron gas (2DEG) confined to a half-plane. Note that Haas et al. first introduced the QHD theory by solving the nonlinear Schr\"{o}rdinger-Poisson or Wigner-Poisson kinetic models\cite{Haas2000,Manfredi2001}. Further, a quantum effect always exists in a metallic electron gas\cite{Haas2011}, establishing new chiral edge modes and enhancing both the splitting and the transverse confinement significantly for a wide range of transverse modes.
Finally, Gauss units will be adopted throughout this paper,
except in the case of specific definitions.

We consider CBPs along the edges of 2DEG while examining the quantum effect without a magnetic field.
The excitations of 2DEG
are described by the QHD model\cite{Haas2000,Manfredi2001} using
the continuity equation
\begin{equation}
\frac{\partial n_{e}}{\partial t}+\nabla\cdot
(n_{e}\mathbf{V})=0,  \label{ne}%
\end{equation}
the momentum-balance equation
\begin{equation}
\frac{\partial \mathbf{u}_{e}}{\partial t}+(\mathbf{u}_{e}\cdot
\nabla) \mathbf{u}_{e}=\frac{e}{m_{e}}\nabla\phi
-\frac{\pi \hbar ^{2}}{m_{e}^{2}}\nabla n_{e}
-\frac{\hbar^{2}}{2m_{e}}\nabla (\frac{1}{\sqrt{n_{e}}}\nabla^{2}\sqrt{n_{e}})
,  \label{mom}
\end{equation}%
and the integral form of the electric potential
\begin{equation}
\phi(\mathbf{r},t)=\int d^{2}\mathbf{r}' W(\mathbf{r}-\mathbf{r}')(n_{e}(\mathbf{r'},t)-n_{0}).
\end{equation}
Here, $m$ is the effective electron mass, $e$ is the elementary charge, $\hbar$ is
the Planck constant, $W(\mathbf{r}-\mathbf{r}')$ is the Coulomb interaction,
$n_{e}$ is the density, and $\mathbf{u}_{e}$ is the homogeneous fluid velocity.
In particular,
\begin{equation}
\mathbf{V}=\mathbf{u}_{e}+\mathbf{V}_{a}, \mathbf{V}_{a}=\frac{eF}{\hbar}[(\nabla\phi)\times \mathbf{e}_{z}],
\end{equation}
with chiral velocity $\mathbf{V}_{a}$, which is a self-induced anomalous
velocity component\cite{Xiao2010}, and $F$ is dimensionless\cite{Song2016}.
Note that the final two terms on the right hand side of Eq. (\ref{mom}) are regarded as quantum effects\cite{Haas2000,Manfredi2001}. That is, the second term is the quantum statistical effect (second-order quantum correction), which is the force due to the internal interactions in the electron species, and the third term is the quantum diffraction effect (fourth-order quantum correction) due to the quantum pressure (the so- called "Bohm potential").
For definiteness, only the quantum statistical effect (second order) is considered in this semi-analytical study, as the inclusion of the quantum pressure term (fourth order) cannot yield an analytical solution.
The above nonlinear equations can be linearized under weak perturbation of the plasmons, where $n_{e}(\mathbf{r},t)$ is represented by the first-order
perturbed value $n_{e}=n_{0}+n_{e1}$, with $n_{e1}\ll n_{0}$\cite{Fetter1985}. $n_{0}$ is the initial equilibrium density.
Thus, the linearized QHD theory yields
\begin{equation}
\frac{\partial n_{e1}}{\partial t}+n_{0}\nabla\cdot
\mathbf{V}=0,  \label{num2d}%
\end{equation}
\begin{equation}
\frac{\partial \mathbf{u}_{e}}{\partial t}=\frac{e}{m}\mathbf{\nabla}\phi
-\frac{\pi \hbar ^{2}}{m^{2}}\nabla n_{e1}
,  \label{mom2d}
\end{equation}%
and
\begin{equation}
\phi(\mathbf{r},t)=\int d^{2}\mathbf{r}' W(\mathbf{r}-\mathbf{r}')n_{e1}(\mathbf{r'},t). \label{phi}
\end{equation}

A semi-infinite ($x\geq0$) 2DEG in the half $x$-$y$ plane is considered, where
 $n_{e}(\mathbf{r},t)$ and $\mathbf{V}(\mathbf{r},t)$ take finite values for  $x\geq0$, while $n_{e}(\mathbf{r},t)=0$ and $\mathbf{V}(\mathbf{r},t)=0$ for $x<0$, as shown in Fig. 1. Thus, an edge plasmon is formed along the boundary at $x=0$, propagating as a plane wave along the $y$ axis with frequency $\omega$ and wave number $q$. The potential, perturbed density, and velocities can be expressed as
\begin{equation}
\begin{array}{lcl}
    \phi(\mathbf{r},t)=\phi_{q}(x)\exp(i\omega t-iqy), \\
    n_{e1}(\mathbf{r},t)=n_{e1q}(x)\exp(i\omega t-iqy), \\
    u_{e}(\mathbf{r},t)=u_{eq}(x)\exp(i\omega t-iqy), \\
    \mathbf{V}(\mathbf{r},t)=\mathbf{V}_{q}(x)\exp(i\omega t-iqy),
 \end{array}
 \label{Fourier}
\end{equation}%
respectively. Thus, Eq. (\ref{num2d}) can be rewritten as $n_{e1}(\mathbf{r})=-n_{0}\mathbf{\nabla}\cdot\mathbf{V}(\mathbf{r},t)/i\omega$. Further, using the step-like property of $\mathbf{V}(\mathbf{r},t)$ [$\mathbf{V}(\mathbf{r},t)\neq0$ for $x\geq0$ and $\mathbf{V}(\mathbf{r},t)=0$ for $x<0$], a jump condition for the $x$ derivative of $\phi_{q}(x)$ can be obtained based on  Eq. (\ref{phi})\cite{Song2016}, where
\begin{equation}
\frac{\partial \phi_{q}}{\partial x}\mid_{x=0^{+}}-\frac{\partial \phi_{q}}{\partial x}\mid_{x=0^{-}}=\frac{1}{i\omega}(\frac{\partial W_{q}}{\partial x}\mid_{x=0^{-}}-\frac{\partial W_{q}}{\partial x}\mid_{x=0^{+}})n_{0}\mathbf{V}_{x}\mid_{x=0^{+}}. \label{refe7}%
\end{equation}
Here, $\mathbf{V}_{x}\mid_{x=0^{+}}=\mathbf{u}_{ex}\mid_{x=0^{+}}+\mathbf{V}_{ax}\mid_{x=0^{+}}$ and $W_{q}=-e\int dk exp(ikx)/\sqrt{q^{2}+k^{2}}$, assuming a wave number $k$ along the $x$ axis.

According to the literature\cite{Song2016,Fetter1985}, $W_{q}$ can be simplified as
\begin{equation}
W_{q}(x)\approx-4\pi e\int\frac{dk}{2\pi}\frac{|q|}{k^{2}+2q^{2}}\exp(ikx)=-\frac{4\pi e}{2\sqrt{2}}\exp(-|q|x), x\geq0,
\end{equation}
for small $ k/q $.
Again, by substituting the simplified $W_{q}(x)$ into the integral potential, we find
\begin{equation}
\phi_{q}(x)=\int dx' W_{q}(x-x')n_{e1q}(x').
\end{equation}
Then, using the convolution theorem, we can obtain a differential
equation of $\phi_{q}(x)$ satisfying
\begin{equation}
\begin{array}{lcl}
 (x^{2}-2q^{2})\phi_{q}(x)=0, x<0, \\
 (x^{2}-2q^{2})\phi_{q}(x)=4\pi e |q|n_{e1q}(x), x\geq0.
 \end{array}
 \label{phi12}
\end{equation}%
Thus, $\phi_{q}(x)$ has solutions of
\begin{equation}
\begin{array}{lcl}
 \phi_{q}(x)=\phi_{1}\exp(\kappa_{1}x), x<0,\\
\phi_{q}(x)=\phi_{2}\exp(-\kappa_{2}x), x\geq0.
 \end{array}
 \label{phi1212}
\end{equation}%
Here, $\kappa_{1}=\sqrt{2}|q|$
and $\kappa_{2}$ is determined as follows.
Applying $\partial/\partial t$ to the continuity relation in Eq. (\ref{num2d}) and
substituting $\frac{\partial \mathbf{V}}{\partial t}=\frac{\partial \mathbf{u_{e}}}{\partial t}+\frac{\partial \mathbf{V_{a}}}{\partial t}$ into Eq. (\ref{num2d}), we find
\begin{equation}
  \frac{\partial^{2}n_{e1}}{\partial t^{2}}=-\frac{e}{m}n_{0}\nabla
^{2}\phi+\frac{\pi \hbar^{2} n_{0}}{m^{2}}\nabla^{2}n_{e1}
. \label{refe4}%
\end{equation}
Using the plane-wave forms of $n_{e1}(\mathbf{r},t)$ and $\phi(\mathbf{r},t)$ and the relation $ n_{e1q}(x)=(x^{2}-2q^{2})\phi_{q}(x)/(4\pi e |q|)$ based on Eq. (\ref{phi12}), $n_{e1}(\mathbf{r},t)$ can be expressed as
\begin{equation}
n_{e1}(\mathbf{r},t)=\frac{\phi_{2}}{4\pi e|q|}(\kappa_{2}^{2}-2q^{2})e^{-\kappa_{2}x}e^{i\omega t-iqy}, \label{ne1q}
\end{equation}
for $x\geq0$.
Therefore, Eq. (\ref{refe4}) can be rewritten as
\begin{equation}
1=\frac{2\omega_{P}^{2}}{\omega^{2}}\frac{\kappa_{2}^{2}-q^{2}}{\kappa_{2}^{2}-2q^{2}}
-A_{Q}\frac{\kappa_{2}^{2}-q^{2}}{\omega^{2}}, \label{kappa2}
\end{equation}
where the bulk plasmon frequency $\omega_{P}=\sqrt{2\pi n_{0}e^{2}|q|/m}$, according to the literature\cite{Fetter1985}. Here, we assume that $q\ll k_{F}=\sqrt{2\pi n_{0}}$, where $k_{F}$ is the Fermi wave number. In particular, $A_{Q}=\pi\hbar^{2}n_{0}/m^{2}$ is the coefficient of the quantum term; therefore, $A_{Q}(\kappa_{2}^{2}-2q^{2})$ characterizes the quantum effect term.
Equation (\ref{kappa2}) yields a nonlinear equation of $\kappa_{2}$. Further, $\kappa_{2}^{2}$ can be explicitly expressed as
\begin{equation}
\kappa_{2}^{2}=\frac{3A_{Q}q^2 - \omega^2 + 2 \omega_{P}^2 \mp
  \sqrt{A_{Q}^2q^4 + 2A_{Q}q^2\omega^2 + \omega^4 + 4A_{Q}q^2\omega_{P}^2 - 4\omega^2\omega_{P}^2 + 4\omega_{P}^4}}{2A_{Q}}. \label{kappa22}
\end{equation}
 At $q=0$, $\kappa_{2}^{2}$ has two solutions: $\kappa_{2}^{2}=0$ and $\kappa_{2}^{2}=-2\omega^{2}$. We consider positive $\kappa_{2}^{2}$ only, as the mode for negative $\kappa_{2}^{2}$ joins the bulk. Thus, the potential decays exponentially from the boundary $x=0$, yielding non-vanishing edge modes.

In what follows, in order to obtain the dispersion relation, we first express
\begin{equation}
{V}_{x}|_{x=0^{+}}={u}_{ex}|_{x=0^{+}}+{V}_{ax}|_{x=0^{+}}.
\end{equation}
Here,
\begin{align}
{u}_{ex}|_{x=0^{+}}=\frac{1}{i\omega}[-\frac{e}{m}\kappa_{2}\phi_{2}
+\frac{\pi\hbar^{2}}{m^{2}}\frac{1}{4\pi e|q|}\kappa_{2}(\kappa_{2}^{2}-2q^{2})\phi_{2}], \label{ue}
\end{align}
based on the Fourier transform of the momentum-balance equation and Eq. (\ref{ne1q}),
and
\begin{equation}
{V}_{ax}|_{x=0^{+}}=\frac{-iqeF}{\hbar}\phi_{2}.
\end{equation}
The dispersion relation is determined from the continuity condition of $\phi$ and the jump boundary condition given in Eq. (\ref{refe7}), such that
\begin{equation}
\sqrt{2}|q|+\kappa_{2}=\frac{2\omega_{p}^{2}}{\omega^{2}}\kappa_{2}-A_{F}\frac{q^{2}sgn(q)}{ \omega}
-A_{Q}\frac{\kappa_{2}(\kappa_{2}^{2}-2q^{2})}{\omega^{2}}, \label{wq}
\end{equation}
indicating a nonlinear equation of $\kappa_{2}$ and $\omega$. Here, $A_{F}=4\pi e^2F/\hbar$ is the coefficient of the chiral Berry term.
At $q=0$, the above dispersion relation yields
$\kappa_{2}=-A_{Q}\kappa_{2}^{3}/\omega^2$, indicating a diverging mode or $\kappa_{2}=0$, which will induce a jump in the edge-mode frequency at $q=0$.
We can couple the two nonlinear equations (Eqs. (\ref{kappa2}) and (\ref{wq})) in order to semi-analytically solve $\kappa_{2}$ and $\omega$ for a given $q$. The solutions obey two coupled polynomial functions of the fifth and fourth orders, respectively.
It is instructive to discuss the properties of Eqs.(\ref{kappa2}) and (\ref{wq}) in the vicinity of $q=0$.
Indeed, the coupling of these two equations yields the solution
\begin{align}
\omega=
&- (2.8A_{Q}q\kappa_{2}^4)/A_{F} - (2.8A_{Q}q\kappa_{2}^3)/A_{F} + (0.5(- 1.4A_{F}^2q^2 + 9.9A_{Q}q^2 + 5.7\omega_{P}^{2})\kappa_{2}^2)/(A_{F}q) \notag \\
&+ (2.9(A_{Q}q^2 + \omega_{P}^{2})\kappa_{2})/(A_{F}q) - (- 0.4A_{F}^2q^2 + 2.1A_{Q}q^2 + 2.1\omega_{P}^{2})/(A_{F}q). \label{wqe}
\end{align}
Here, $\omega$ represents $\omega_{+edge}$ and $\omega_{-edge}$ for negative and positive $q$, respectively.

Thus, we can obtain the limits of both $\omega$ and $\kappa_{2}$ for $q\rightarrow 0^{-}$ and $q\rightarrow 0^{+}$:
\begin{equation}
\begin{array}{lcl}
 \kappa_{2}|q\rightarrow 0^{-}=O(q^{4}),   \\
\omega|q\rightarrow 0^{-}=\frac{21\omega_{p}^2}{10A_{F}|q|},  \\
 \kappa_{2}|q\rightarrow 0^{+}=\frac{\sqrt{2}}{2}q,  \\
\omega|q\rightarrow 0^{+}=\frac{\sqrt{2}}{8}A_{F}q.
 \end{array}
 \label{limits}
\end{equation}%
The results presented below are determined by taking
$n_{0}=6\times10^{10}$ cm$^{-2}$ and $m=0.03m_{e}$, based on the literature\cite{Yan2012,Song2016}, where $m_{e}$ is the free electron mass.
For definiteness, we take the unit of $q$ as $k_{F}$ for large $q$, but as cm$^{-1}$ for small $q$, and the unit of the plasmon frequency $\omega$ is $E_{F}/\hbar$ with the Fermi energy $E_{F}=\hbar^{2}k_{F}^2/2m$. Thus, the strength of the quantum effect is defined by a dimensionless parameter $\widetilde{Q}=A_{Q}(\hbar k_{F}/E_{F})^2$.

Figures 2(a, b) show the dispersion relations of CBPs with $F=1$, comparing the edge modes obtained without (circles) and with (triangles) the quantum effect.
The $\omega_{+ edge}$ mode without the quantum effect diverges between $- 0.0028<q/k_{F}<-0.0013$, and both $\omega_{\pm edge}$ modes with the quantum effect diverge between $0.0024<|q|/k_{F}\leq0.0033$ and $q=0$.
Indeed, with the quantum effect, substituting these $q$ values into
Eqs. (\ref{kappa2}) and (\ref{wq}) yields negative decay constant $\kappa_{2}$ and edge frequencies $\omega_{\pm edge}$, inducing the diverging modes.
Without the quantum effect, in order to validate our model, we first compare our results with those of the literature\cite{Song2016} for small $q$. The splitting trends of the edge modes agree well with the linear approximation results (the first-order $q$ modes) given in Ref.\cite{Song2016}, as shown in Fig. 2(a) (circles and squares ($\omega_{P}$)).
For large amplitude and negative $q$, a new edge mode emerges for $|q|/k_{F}>0.0028$ (circles) (Fig. 2(a)), due to the dominance of the quadratic ($\propto q^2$) terms; this behavior did not occur for the linear approximation limit.
With the quantum effect,
$\omega_{+edge}$ exhibits a maximum value in the vicinity of $q=0^{-}$,  and decreases with increasing $|q|$ for $|q|/k_{F}\leq0.0024$ (Fig. 2(a)). This behavior occurs because the dispersion relation with the quantum effect has a vanishing solution at $q=0$ (see the description below Eq. (\ref{wq})).
This scenario is reversed for the $\omega_{-edge}$ mode, where $\omega_{-edge}$ exhibits minimum values near $q=0^{+}$ and increases quickly with increasing $|q|$, for large $q$.

Figure 2(b) indicates that $\omega_{\pm edge}$, respectively, exhibits a new one-way edge state with the quantum effect compared to that without the quantum effect, at the both side of $q=0$.
Although this one-way edge CBP state discussed here presents some similar characters with
the one-way edge magnetoplasmon state reported in Ref.\cite{Jin2016}, there are fundamental differences between them. The one-way edge magnetoplasmon has been realized by breaking the time-reversal symmetry with oppositely-directed magnetic fields, whereas the one-way edge CBP is realized by considering the quantum effect.
Indeed, the $\omega_{\pm edge}$ with the quantum effect have different limits (see Eq. (\ref{limits})) for $q\rightarrow 0^{-}$ and $q\rightarrow 0^{+}$, whereas the $\omega_{\pm edge}$ without the quantum effect have symmetric distribution at the different sides of $q=0$ and tend to zero at $q=0$ (see (Inset) in 2(b)).
Note that the splittings have crucial applications in subwavelength optical
nonreciprocity\cite{Sounas2013} in the terahertz (THz) range.

The corresponding transverse confinements of these edge modes are displayed in Fig. 3 , for (a) negative and (b) positive $q$, for the same conditions as in Fig. 2. Without the quantum effect, Fig. 3(a) again shows divergence of $\omega_{+edge}$  for $- 0.0028<q/k_{F}<-0.0013$, corresponding to Fig. 2(a) (circles).
With the quantum effect, both $\omega_{\pm edge}$ modes exhibit greater localization over a wider range of $q$; this behavior is ascribed to the coupling of the quantum effect and the chiral velocity (see Eq. (\ref{wqe})). This significantly enhanced transverse localization is preferable for the realization of subwavelength optical non-reciprocal devices in the THz range with no magnetic field. Note that this possibility avoids the current requirement for large magnetic-based devices to realize nonreciprocity in the THz range, enhancing industrial potential as regards microwave and photonic components.

Without the quantum effect, the splitting between $\omega_{\pm edge}$  increases linearly with increasing $q$ and $F$, because of the relations $\omega \propto Fq$ and $\mathbf{V}_{a}=\frac{eF}{\hbar}[(\nabla\phi)\times \mathbf{e}_{z}]$, which are given in Ref.\cite{Song2016}. However, with the quantum effect, the splitting  between $\omega_{\pm edge}$ indicates a nonmonotonous relation to $q$ and $F$, because of the diverging solutions at certain $q$  and $F$ values. Indeed, diverging solutions usually occur for a higher-order polynomial function. Fig. 4  shows comparisons of (a) the edge modes $\omega_{\pm edge}$ and (b) the corresponding transverse confinement lengths, with and without the quantum effect, as a function of $F$, and for fixed $q/k_{F}=0.0098$.
Without the quantum effect,
in Fig. 4(a), the splitting modes (squares and spheres) agree well with those in the literature\cite{Song2016}. However, a new $\omega_{+ edge}$ mode branch emerges as a result of the quadratic $q^{2}$ terms for $F\geq 0.55$; this new mode was easily omitted from the linear $q$ approximation. As a result, this new mode introduces a new confinement length branch for $F\geq 0.55$, as can be clearly seen in Fig. 4(b) (squares). In contrast, the confinement lengths of the $\omega_{- edge}$ mode and the increasing branch of the $\omega_{+ edge}$ are in good agreement with the curves in Ref.\cite{Song2016}. This increasing branch diverges when $F\geq 0.35$, and a new $\omega_{+ edge}$ mode branch develops when $F\geq 0.55$. The latter is more closely compressed to the edge than its counterpropagating $\omega_{- edge}$ mode.
With the quantum effect, $\omega_{+ edge}$ vanishes for $F< 0.55$. This behavior occurs because small $F$ yields negative decay constants and vanishing frequencies. Again, the oscillations in the $\omega_{- edge}$ mode can be attributed to the fact that the dispersion relation is determined by polynomial functions of the high-order decay constant $\kappa_{2}$, where the coefficients change non-monotonically with increasing $F$,  as shown in Eq.(\ref{wqe}). With increasing $F$, the $\omega_{+ edge}$ mode introduces a cut-off confinement of 0.4 (upright triangles), and this mode is primarily confined to the edge in both cases, i.e., both with and without the quantum effect. Such a confinement will have crucial importance for the realization of nonreciprocal devices with no magnetic field.

Finally, we briefly discuss the properties of CBPs with the quantum effect, compared to magnetoplasmons.
For magnetoplasmons, the authors of Ref.\cite{Fetter1985} have reported that the quantum effect with strength defined as being $\propto sq$ has almost no effect on the edge modes in the limit $sq<<\omega_{P}$, with $s^2=m^{-1}(\partial p/\partial n)$. This means that the edge magnetographs are almost independent of the quantum effect.
In contrast, for the CBPs, the quantum effect induces several new edge modes (see the $\omega_{\pm edge}$ obtained with the quantum effect in the above figures), even for the limit $sq<<\omega_{P}$ (defined as $A_{Q}q^2<<\omega_{P}^2$ in the present work). Indeed, the quantum effect clearly enhances the chirality as well as the transverse confinement, indicating strong coupling of both these properties (see Eq. (\ref{wqe})).
The chirality in CBPs emerges from the coupling action of the self-induced electric field $-\nabla \phi$, due to the collective density oscillation and the anomalous velocity $\mathbf{V}_{a}=\frac{eF}{\hbar}[(\nabla\phi)\times \mathbf{e}_{z}]$, the strength of which depends on $-\nabla \phi$  and $F$. $\mathbf{V}_{a}$ can be realized by $\Omega_{n}(\mathbf{k})$ , which is an intrinsic quantum-mechanical property of a perfect crystal, because it depends on the Bloch wave function only.
Indeed, $\mathbf{V}_{a}$ can be expressed in terms of $\Omega_{n}(\mathbf{k})$ in a given Bloch state $u_{n}(\mathbf{k})$ in the $n^{th}$ band\cite{Xiao2010}, such that
\begin{equation}
\mathbf{V}_{a}=\frac{e}{\hbar} \nabla\phi\times\Omega_{n}(\mathbf{k}), \Omega_{n}(\mathbf{k})=i<\nabla_{\mathbf{k}} u_{n}(\mathbf{k})|\times|\nabla_{\mathbf{k}} u_{n}(\mathbf{k})>. \label{BC}
\end{equation}

In summary, we used a quantum hydrodynamic theory to study the chiral Berry plasmons in a semi-infinite 2DEG, with the chiral Berry term and the quantum term being considered simultaneously.
Note that this semi-infinite 2DEG can be realized easily in experiment using 2D gapped Dirac
materials, such as graphene and transition metal dichalcogenide monolayers\cite{Xiao2007,Xiao2012}.
Herein, the transverse decay constant and the edge plasmon frequency were obtained by nonlinearly and semi-analytically solving two coupled transcendental equations. First, our model was validated by comparing our results with those based on a linear approximation method with no quantum effect, for small $q$.
In our calculation, without the quantum effect, a new chiral edge mode emerged, which was due to the quadratic term contribution ($\propto q^2$).
With the quantum effect, our results predicted that both the chiral edge modes and the transverse confinement in a bounded 2DEG can be tuned. It was found that the quantum effect introduces new and intriguing phenomena. That is, a new one-way chiral edge state with the quantum effect compared to that without the quantum effect, at the both side of $q=0$ was found. Indeed, the $\omega_{\pm edge}$ modes exhibit different limits for $q\rightarrow 0^{-}$ and $q\rightarrow 0^{+}$, in contrast with the case without the quantum effect and $\omega_{P}$.
The newly discovered one-way chiral edge modes will elucidate new
paradigms in condensed matter and plasma-optics physics\cite{Sounas2013,Barnes2003,Yao2008} for the study of bounded 2DEGs, such as graphene and traditional quantum dots.

This work was supported by NSFC (11405067, 11375163, 11575135).

\bibliography{201712TIC}

\begin{center}
\textbf{Figures and figure captions}
\end{center}

\begin{figure}[tbph]
\begin{center}
\includegraphics[height=4.0cm,width=6.0cm]{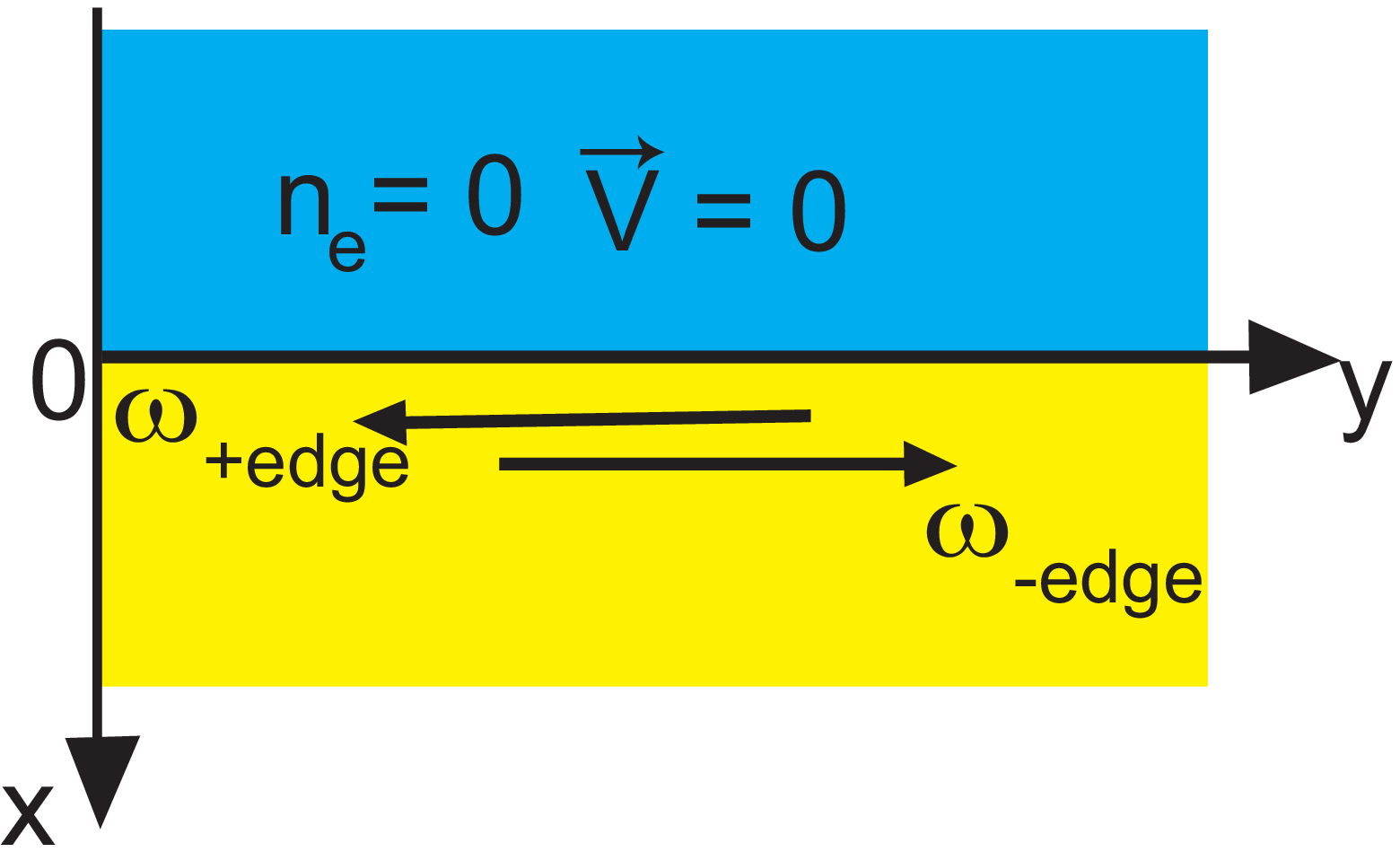}
\end{center}
\caption{(Color online) Schematic illustration of metal and vacuum interface. Chiral Berry plasmons (CBPs) are realized along the edges of a two-dimensional metal surface with a finite Berry flux.}
\label{fig:Fig1}
\end{figure}

\begin{figure}[tbph]
\begin{center}
\includegraphics[height=18.0cm,width=14.0cm]{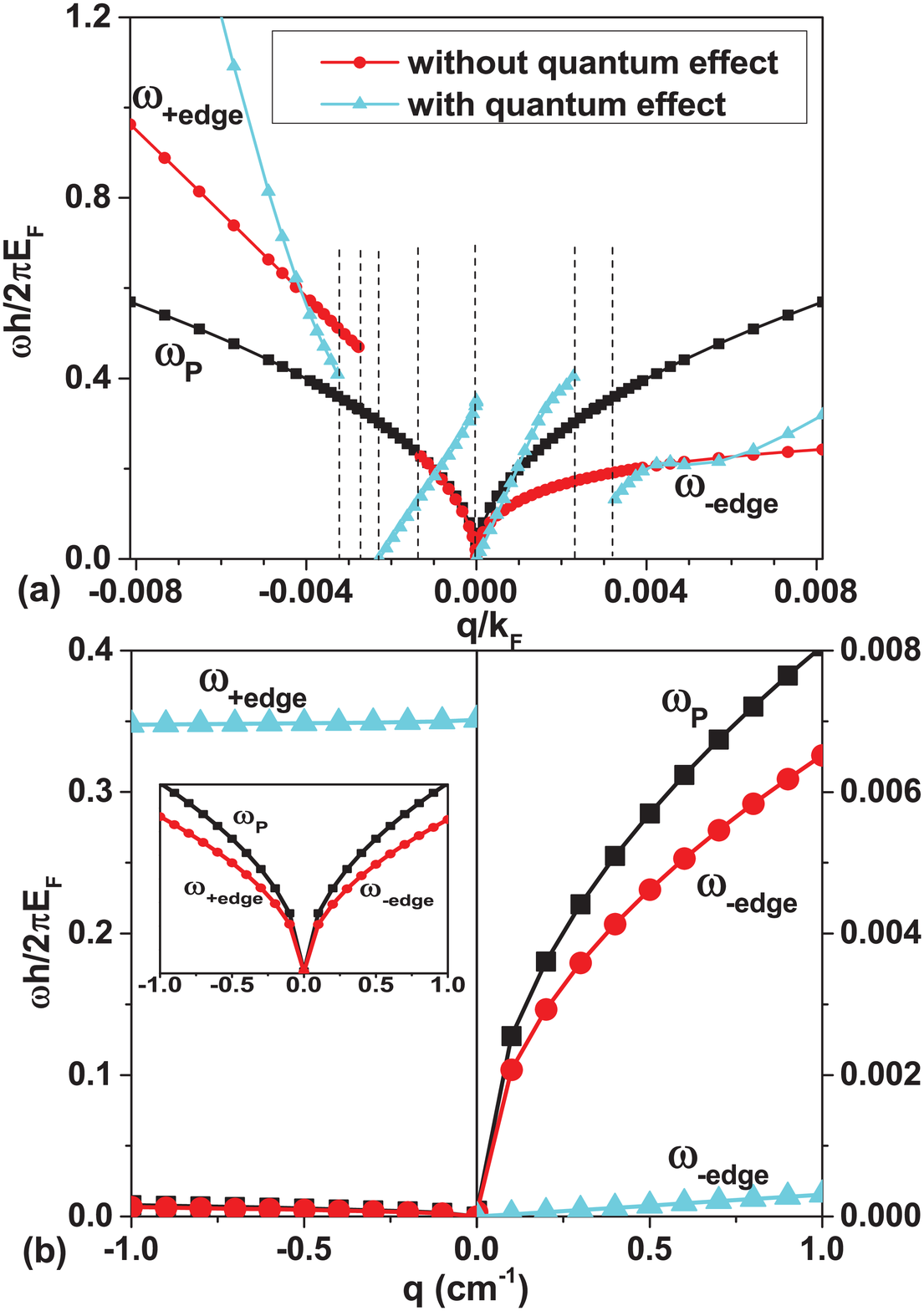}
\end{center}
\caption{(Color online) (a) CBP dispersion relations without (circles) and with (triangles) quantum effect, compared to bulk plasmon frequency $\omega_{P}$ (squares). (b) Magnified view of part of (a) showing enlarged curves near $q=0$, where a new one-way edge state is found at the both sides of $q=0$ with the quantum effect. (Inset) in (b) The comparison between $\omega_{P}$ and $\omega_{\pm edge}$ without the quantum effect, illustrating the symmetry of the $\omega_{\pm edge}$ in the vicinity of $q=0$ without the quantum effect.
A larger deviation of $\omega_{\pm edge}$ from $\omega_{P}$ is observed with the quantum effect, indicating stronger chirality. Note that the $\omega_{+ edge}$ mode without the quantum effect diverges between $- 0.0028<q/k_{F}<-0.0013$, whereas both $\omega_{\pm edge}$ modes with the quantum effect diverge between $0.0024<|q|/k_{F}\leq0.0033$ and $q=0$. Here, $F=1$. }
\label{fig:Fig2}
\end{figure}

\begin{figure}[tbph]
\begin{center}
\includegraphics[height=8.0cm,width=16.0cm]{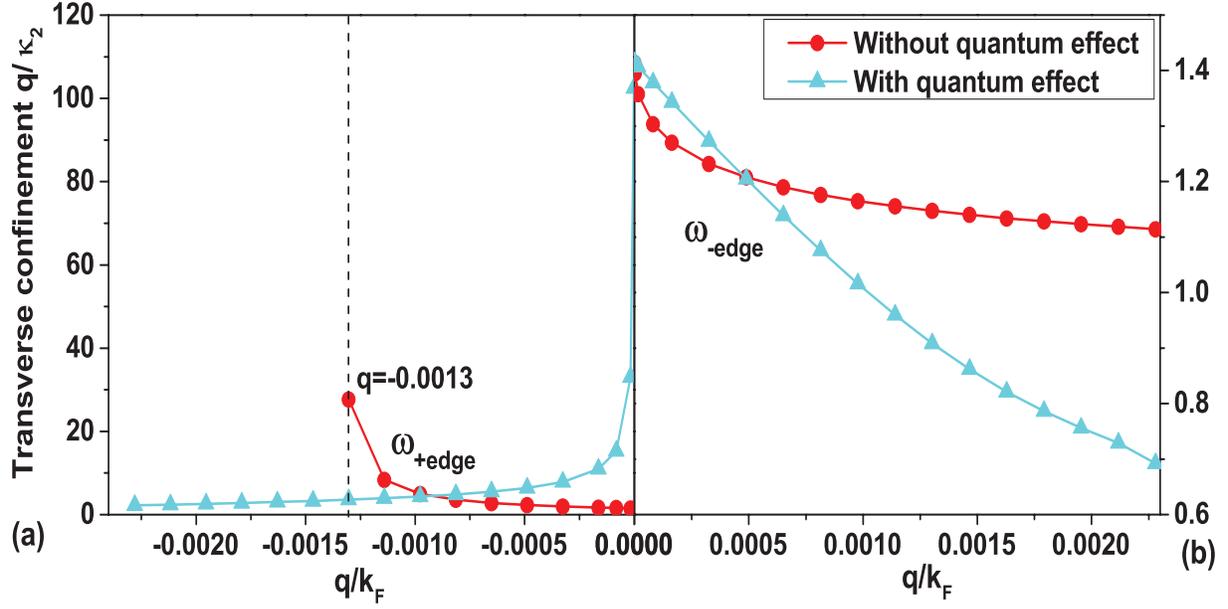}
\end{center}
\caption{(Color online) Transverse confinement lengths of CBPs with $F=1$ expressed as $q/\kappa_{2}$ for (a) $\omega_{+edge}$ and (b) $\omega_{-edge}$ modes, without (circles) and with (triangles) quantum effect. With the quantum effect, both $\omega_{\pm edge}$ modes become more closely restricted to the edge under increasing $q$.  }
\label{fig:Fig3}
\end{figure}

\begin{figure}[tbph]
\begin{center}
\includegraphics[height=16.0cm,width=10.0cm]{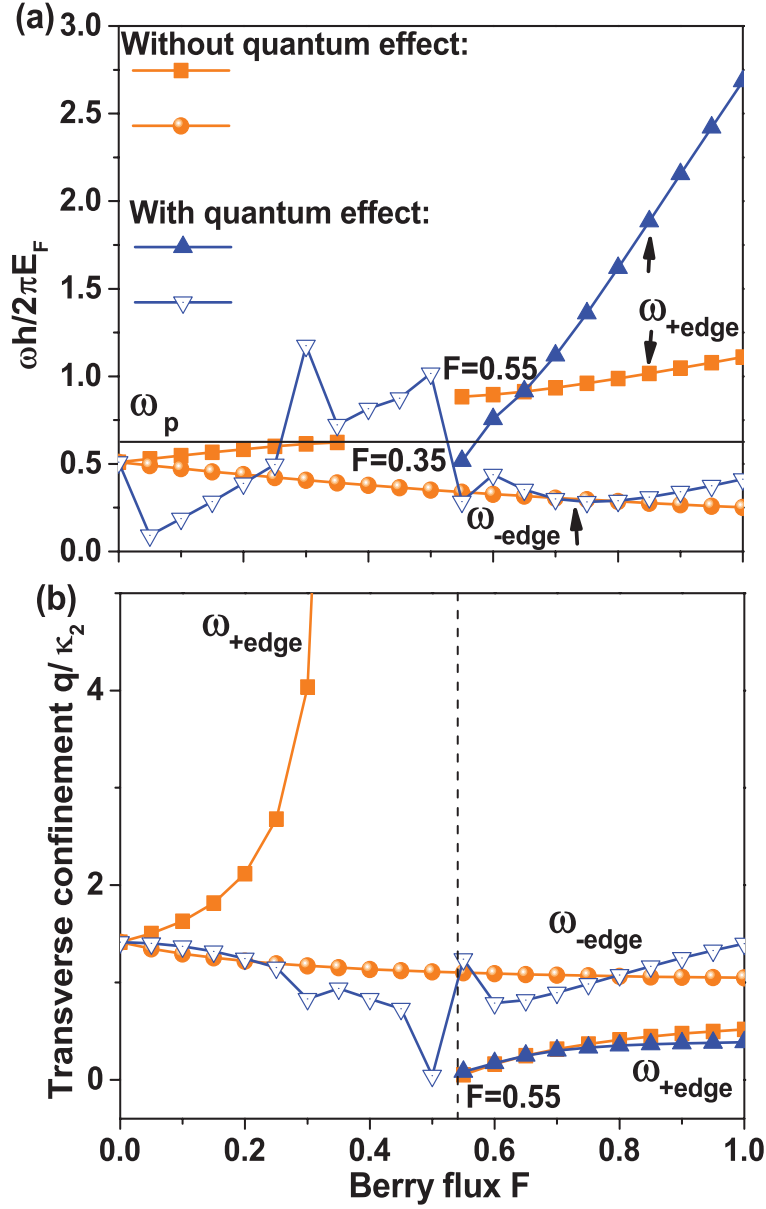}
\end{center}
\caption{(Color online) (a) Frequencies and (b) transverse confinement lengths of CBPs as functions of $F$, without ($\omega_{+edge}$: squares, $\omega_{-edge}$: spheres) and with ($\omega_{+edge}$: upright triangles, $\omega_{-edge}$: inverted triangles) quantum effect.
The transverse confinement length is expressed as $q/\kappa_{2}$. The frequency splitting becomes larger with the quantum effect. With increasing $F$, the $\omega_{+edge}$ mode with the quantum effect indicates a cut-off confinement of 0.4, and this mode is primarily confined to the edge. Note that the $\omega_{+edge}$ mode without the quantum effect diverges for $F=0.35$-$0.55$, while the $\omega_{+edge}$ mode with the quantum effect diverges for small $F<0.55$, where no point is shown. Here, $q/k_{F}=0.0098$.}
\label{fig:Fig4}
\end{figure}

\end{document}